\documentclass[pra,twocolumn,amsmath,amssymb]{revtex4}
\newcommand{\be}{\begin{equation}}
\newcommand{\ee}{\end{equation}}
\newcommand{\ba}{\begin{eqnarray}}
\newcommand{\ea}{\end{eqnarray}}
\newcommand{\nn}{\nonumber \\}

\usepackage{graphicx}
\usepackage{bm}

\begin{document}

\title{Rabi Oscillations in Systems with Small Anharmonicity}
\author{M.~H.~S.~Amin}
\affiliation{D-Wave Systems Inc., 320-1985 W. Broadway,
Vancouver, B.C., V6J 4Y3 Canada}

\begin{abstract}

When a two-level quantum system is irradiated with a microwave
signal, in resonance with the energy difference between the
levels, it starts Rabi oscillation between those states. If there
are other states close, in energy, to the first two, the Rabi
signal will also induce transition to those. Here, we study the
probability of transition to the third state, in a three-level
system, while a Rabi oscillation between the first two states is
performed. We investigate the effect of pulse shaping on the
probability and suggest methods for optimizing pulse shapes to
reduce transition probability.

\end{abstract}


\maketitle

\section{Introduction}

Most qubits (i.e.~basic elements in a quantum computer) are not
true two-level systems. Yet, only the first two energy states are
commonly considered relevant for quantum computation. As a result,
any transition to the upper levels during the gate operations is a
leakage of information outside the computational space, and
therefore a source of error.

One of the common methods to perform gate operations in a qubit is
via Rabi oscillations \cite{Rabi}. The speed of operation is
determined by the Rabi frequency $\Omega_R$, which is proportional
to the amplitude of the applied microwave signal. Rabi
oscillations have been observed in many quantum systems, including
superconducting qubits
\cite{NECRabi,vion,martinis,chiorescu,ilichev}, excitons in single
quantum dots \cite{QD1,QD2}, and very recently single electron
spins in nitrogen-vacancy defect centers in diamond
\cite{jelezko}.

In a multi-level quantum system, Rabi oscillations may not be
limited to only the first two states. For example, in a harmonic
oscillator, with equally spaced energy eigenvalues, applying a
Rabi signal in resonance with the level spacings will occupy many
states. When the system is strongly anharmonic, on the other hand,
i.e. when the third state is far above the first two, the
probability of transition will be vanishingly small.

To have a quantitative measure of anharmonicity, we define an
anharmonicity coefficient by
\be
 \delta = {(E_{21}-E_{10}) / E_{10}},
\ee
where, $E_{ij}=E_i-E_j$, with $E_0$ being the ground state and
$E_{i>0}$, the $i$-th excited state energy. $\delta$ is zero for a
harmonic oscillator and ${\to}+\infty$ for an ideal two level
system.

Not every qubit realization has large $\delta$. For example, in a
current biased Josephson junction qubit \cite{martinis}, $E_{21}$
is always smaller than $E_{10}$ leading to a negative $\delta$
close to zero. Charge-phase qubits also suffer from small
anharmonicity, merely because of operating in the charge-phase
regime; for the ``quantronium'' qubit of Vion {\em et
al}.~\cite{vion}, $\delta \approx 0.2$, and for the flux based
charge-phase qubit of Ref.~\cite{amin}, a $\delta=O(1)$ was
suggested.

The purpose of this paper is to study how much smallness of
$\delta$ can affect transition to the upper state, and how it can
be prevented. We study the problem in a three-state quantum system
with small anharmonicity. In Sec.~II, we perform analytical
calculations using Rotating Wave Approximation (RWA). Section III,
goes beyond RWA using numerical methods. The effect of pulse
shaping on the transition probabilities is addressed in Sec.~IV.
Section V, discusses practical examples within superconducting
qubit implementations. A brief summary together with some
concluding remarks are provided in Sec.~VI.

\section{Analytical calculation}

Let us consider a quantum system with three states $|i\rangle$,
$i=a,b,c$, irradiated with a microwave signal in resonance with
the energy difference between the first two levels. The
Hamiltonian of the system is written as ($E_c>E_b>E_a=0$)
\be
 H= E_b|b\rangle \langle b| + E_c|c\rangle \langle c| + V(t)
 \label{H}
\ee
where $V(t)=V_0e^{-i\omega_0 t} +$ c.c. is the microwave signal
($\hbar=1$). Writing the wave function as $\psi(t) =
c_a(t)|a\rangle + c_b(t)|b\rangle + c_c(t)|c\rangle$, the
equations of motion for $c_i$ are
\ba
 i\dot c_a &=& V_{ab} c_b, \nn
 i\dot c_b &=& V_{ab}^* c_a + E_b c_b + V_{bc} c_c, \nn
 i\dot c_c &=& V_{bc}^* c_b + E_c c_c,
\ea
where $V_{ij}(t) = \langle i| V(t) |j\rangle$. We have taken
$V_{ac}=0$; the transition probability will be small anyway
because of large frequency difference. For simplicity, we write
$E_b=\omega_0$ and $E_c=(2+\delta)\omega_0$. In this section, we
assume $\delta \ll 1$ to ensure small anharmonicity.

Let us define $\tilde{c}_b = c_b e^{i\omega_0t}$, $\tilde{c}_c =
c_c e^{i2\omega_0t}$, and write $V_{ab}/\omega_0=ue^{-i\omega_0t}
+$ c.c. and $V_{bc}/\omega_0=ve^{-i\omega_0t} +$ c.c. Using RWA,
i.e.~ignoring the fast oscillating terms, we find
\ba
 \partial_\tau {\tilde{c}}_a &=& -iu \tilde{c}_b, \nn
 \partial_\tau {\tilde{c}}_b &=& -iu^* \tilde{c}_a -iv \tilde{c}_c, \nn
 \partial_\tau {\tilde{c}}_c &=& -iv^* \tilde{c}_b -i\delta
 \tilde{c}_c,
 \label{eqm}
\ea
where $\tau=\omega_0t$. The equation of motion for $\tilde{c}_b$
can be extracted from (\ref{eqm}):
\be
 \left[ \partial_\tau^3 + i\delta \partial_\tau^2 + (|u|^2+|v|^2)
 \partial_\tau + i\delta |u|^2 \right] \tilde{c}_b=0.
\ee
Writing $\tilde{c}_b=k e^{-ix\tau}$, $x$ needs to satisfy
\be
 x^3 - \delta x^2 - (|u|^2+|v|^2)x + \delta |u|^2 =0. \label{eqx}
\ee
General solutions are
\be
 x_n = {1\over 3} \left\{ \delta + 2z \cos \left[ \theta +
 (2n-1){\pi\over 3} \right] \right\}, \quad n=1,2,3 \label{xn}
\ee
where
\ba
 z &=& \sqrt{3(|u|^2+|v|^2)+\delta^2} \label{z}, \\
 \theta &=& {1\over 3} \arccos \left({9\delta \left(|u|^2-|v|^2/2 \right)
 -\delta^3 \over z^3}
 \right). \label{theta}
\ea

To find the coefficients, let us write
\ba
 \tilde{c}_a &=& \sum_{n=1}^3 k_n e^{-ix_n\tau}, \nn
 \tilde{c}_b &=& {1\over u} \sum_{n=1}^3 x_n k_n e^{-ix_n\tau}, \nn
 \tilde{c}_c &=& {1\over uv} \sum_{n=1}^3 (x_n^2 - |u|^2) k_n
 e^{-ix_n\tau},
\ea
which satisfy (\ref{eqm}). Assuming that the system starts from
the ground state, we impose the initial conditions:
$\tilde{c}_a=1$ and $\tilde{c}_b=\tilde{c}_c=0$, which yield
\ba
 \sum_{n=1}^3 k_n = 0 , \quad
 \sum_{n=1}^3 x_n k_n = 0 , \quad
 \sum_{n=1}^3 x_n^2  k_n = |u|^2
\ea
Solving these equations for $k_n$, we find
\ba
 k_1 = {|u|^2 + x_2x_3 \over (x_1-x_2)(x_1-x_3)}. \label{k1}
\ea
$k_2$ and $k_3$ can be obtained using the permutation
$1{\to}2{\to}3{\to}1$.

Let us write $\tilde{c}_c=\sum \alpha_n e^{-ix_n\tau}$, where
$\alpha_n = (x_n^2-|u|^2)k_n/uv$. The probability of finding the
system in the upper state is
\be
 P_c(\tau) = |\tilde{c}_c|^2 = \left|\sum_{n=1}^3 \alpha_n e^{-ix_n\tau} \right|^2
 \leq P_{\rm max}
\ee
where
\be
 P_{\rm max}=  \left(\sum_{n=1}^3 |\alpha_n|\right)^2
 \label{EqPmax}
\ee
determines an upper bound for $P_c(\tau)$. We first study the
solution is some special cases.

\subsection{Case I, $\delta=0$}

This is the simplest case that the problem can be solved. From
(\ref{xn})--(\ref{theta}), we find
\be
 x_1=0, \qquad x_{2,3}=\mp \Omega_R,
\ee
where $\Omega_R=\sqrt{|u|^2+|v|^2}$. These can also be found
easily from (\ref{eqx}) directly. Using (\ref{k1}), we find
$k_1=|v/\Omega_R|^2$ and $k_2=k_3=(1/2)|u/\Omega_R|^2$. As a
result
\ba
 \tilde{c}_a &=& {1\over \Omega_R^2}(|v|^2+ |u|^2 \cos \Omega_R\tau), \nn
 \tilde{c}_b &=& -i{u^* \over \Omega_R} \sin \Omega_R\tau, \nn
 \tilde{c}_c &=& -{u^*v^* \over \Omega_R^2}(1-\cos \Omega_R\tau).
\ea
The system oscillates with only one frequency $\Omega_R$. The
probability of finding the system in the upper state
$|\tilde{c}_c|^2$ can become large: $P_{\rm max}=
4|uv|^2/\Omega_R^4$. This is expected in a system with zero
anharmonicity.

\subsection{Case II, $v=0$} Using (\ref{z})--(\ref{theta}), together with
%
%
\be
 \cos 3\theta = 4 \cos^3 \theta - 3\cos \theta,
\ee
we find $\cos \theta = -\delta/z$, which immediately gives
\ba
 x_1=-|u|, \qquad x_2=\delta, \qquad x_3=|u|.
\ea
These could also be found directly from (\ref{eqx}). For $k$'s, we
get: $k_1=k_3=1/2$ and $k_2=0$, leading to
\ba
 \tilde{c}_a &=&  \cos \Omega_R \tau, \nn
 \tilde{c}_b &=&  -i{\Omega_R\over u} \sin \Omega_R \tau, \nn
 \tilde{c}_c &=&  0.
\ea
The results show usual Rabi oscillation between the first two
states with frequency $\Omega_R=|u|$. The probability of finding
the system in the upper state is always zero ($P_c=0$), as
expected because $v=0$.

\subsection{Case III, $\delta \gg u,v$}

In the regime $u,v \ll \delta \ll 1$, one can find asymptotic
solutions. A systematic expansion in $u/\delta$ and $v/\delta$
gives
\ba
 x_1 &=&  |u| \left(1 - {|v|^2 \over 2\delta^2} \right) - {|v|^2 \over
 2\delta}\nn
 x_2 &=& -|u| \left(1 - {|v|^2 \over 2\delta^2} \right) - {|v|^2 \over
 2\delta}\nn
 x_3 &=& \delta \left(1 + {|v|^2 \over \delta^2} \right)
\ea
Leading to the Rabi frequency
\be
 \Omega_R = |u| \left(1 - {|v|^2 \over 2\delta^2} \right)
\ee
The dependence of the Rabi frequency on the amplitude of the
microwave signal now has the form
\be
 \Omega_R \propto V_0 \left(1 - \beta V_0^2 \right),
\ee
where the coefficient $\beta$ depends on the details of the
system. The deviation from the proportionality relation is a
signature of transition to the upper states. Such a deviation has
been experimentally observed recently in a current biased dc-SQUID
structure~\cite{claudon}.

The probability of finding the system in the upper state is given
by
\be
 P_c \approx {|v|^2 \over \delta^2} \sin^2 \Omega_R \tau.
\ee
It oscillates with the Rabi frequency $\Omega_R$. The maximum
probability
\be
 P_{\rm max} \approx {|v|^2 \over \delta^2} \approx
 \gamma \left({\Omega_R \over \delta} \right)^2  \label{Panal}
\ee
%
%
%
occurs at half a Rabi period $\tau=\pi/\Omega_R$, where $P_b$ is
the largest. This is not the case for small $\delta$ (see
e.g.~case I). Here, $\gamma=|v/u|^2$ is a constant depending on
the details of the Hamiltonian. In most physical systems $|v|\sim
|u|$ and therefore $\gamma = O(1)$.

\subsection{General case}

\begin{figure}[t]
\includegraphics[width=6cm]{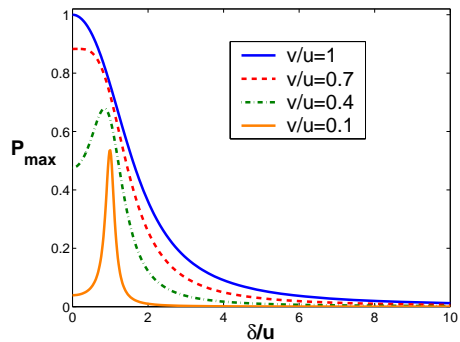}
\caption{$P_{\rm max}$ vs $\delta/u$ for different values of
$v/u$. The curves are symmetric with respect to $\delta \to
-\delta$. }\label{fig1}
\end{figure}

It is not straightforward to find a closed analytical solution for
the general case. Instead we plot the results for $P_{\rm max}$,
calculated using (\ref{xn})--(\ref{theta}) together with
(\ref{k1}) and (\ref{EqPmax}). Figure \ref{fig1} shows $P_{\rm
max}$ as a function of $\delta/u$ with different values of $v/u$.
At small $v/u$, the curves are peaked near $\delta=u$, while for
larger $v/u$ the peak appears near $\delta=0$. In all cases
$P_{\rm max}$ becomes very small at large $\delta/u$, as expected.

\section{Numerical calculation}

\begin{figure}[t]
\includegraphics[width=6cm]{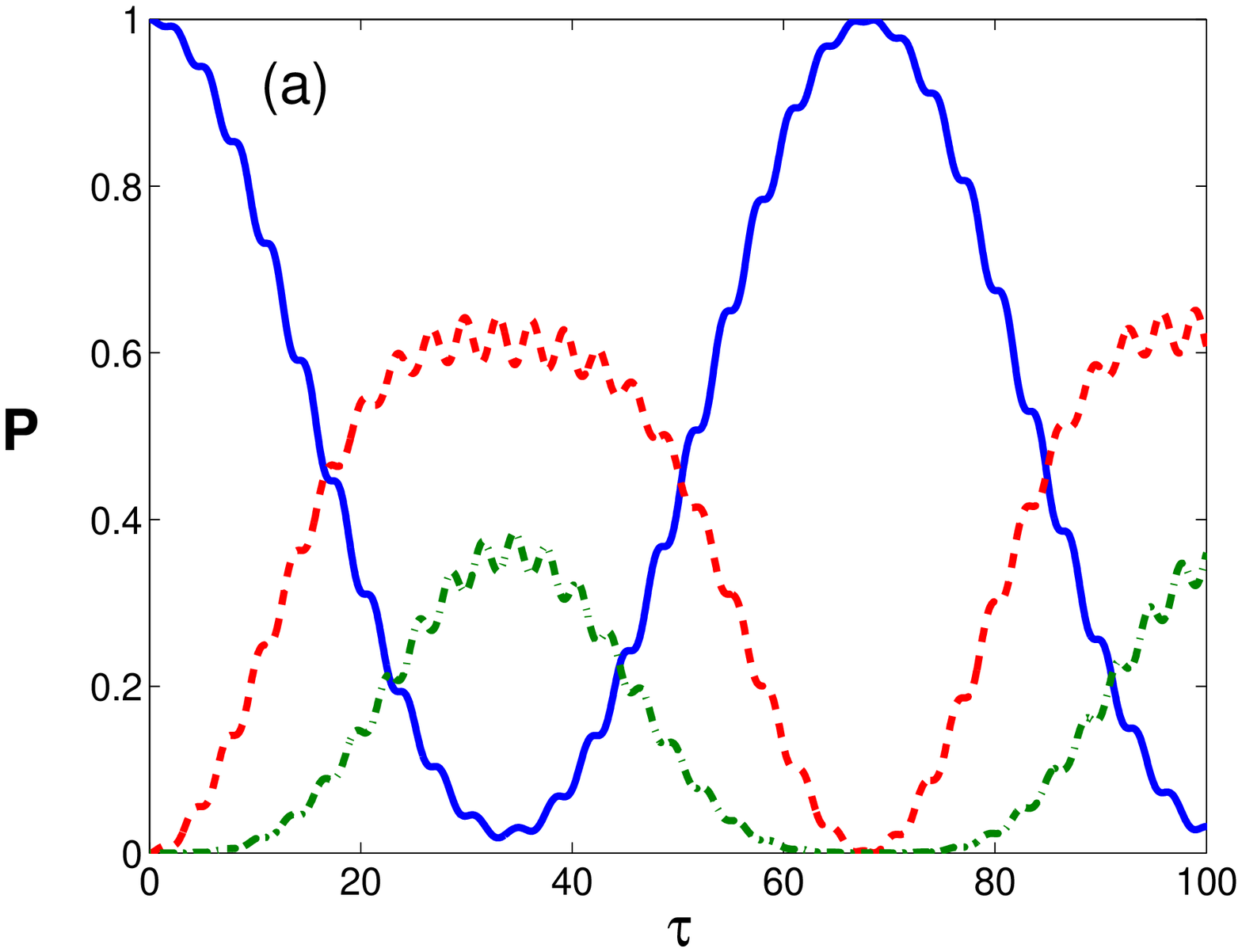}
\includegraphics[width=6cm]{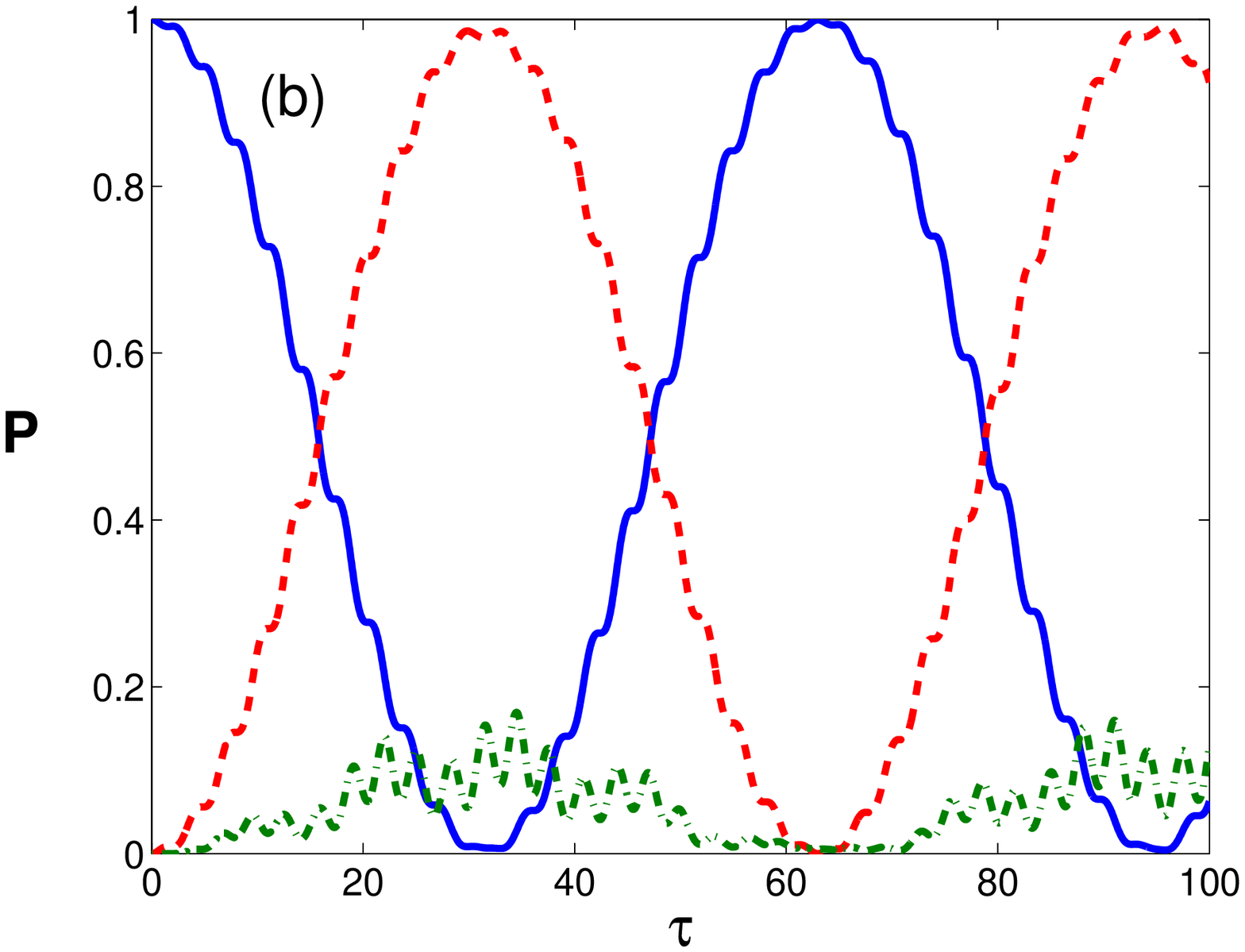}
\caption{Probability $P_a$ (solid), $P_b=\rho_{22}$ (dashed), and
$P_c=\rho_{33}$ (dot-dashed), as a function of time. The
parameters are $u=v=0.1$, $\delta=0.1$ (a) and $\delta=0.5$ (b).
For clarity, $P_c$ in (b) is magnified by a factor of
10.}\label{Rabi}
\end{figure}

In this section we calculate the quantum evolution of the system
numerically using density matrix approach. This allows us to study
the system beyond RWA and/or at large $\delta$. The dynamics of
the $3\times3$ density matrix $\rho$ is described by
\be
 i{d\rho \over dt} = [H,\rho].
\ee
We integrate this equation starting from
\be
 \rho_0 = \left(
 \begin{array}{ccc}
1 & 0 &  0 \\
0 & 0 &  0 \\
0 & 0 &  0
\end{array} \right),
\ee
which describes the system at the lowest energy state.
Probabilities of finding the system in different states are given
by: $P_a=\rho_{11}$, $P_b=\rho_{22}$, and $P_c=\rho_{33}$. Figure
\ref{Rabi} displays the time evolution of these probabilities. The
fast oscillations are the effect of high frequency terms, which
were ignored in the previous section due to RWA. Figure
\ref{Rabi}a shows the Rabi oscillation when $\delta=0.1$. After
(almost) half a Rabi period, significant amount of the probability
goes to the third state. By increasing $\delta$ to 0.5, the
probability of finding the system in the upper state is
significantly reduced (Fig.~\ref{Rabi}b; the curve in the figure
is magnified for clarity).

\begin{figure}[h]
\includegraphics[width=6cm]{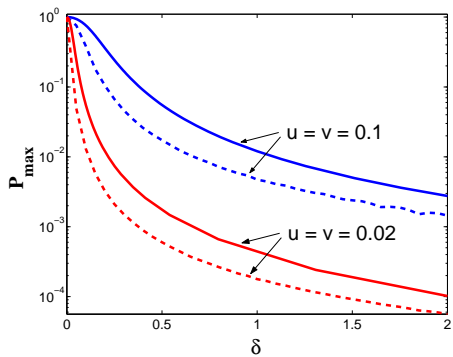}
\caption{$P_{\rm max}$ vs $\delta$ for different values of $u$ and
$v$. Solid (dashed) curves are analytical (numerical) results.
}\label{Pmax}
\end{figure}

\begin{figure}[h]
\includegraphics[width=6cm]{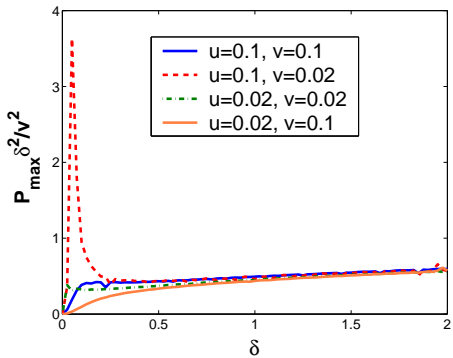}
\caption{$P_{\rm max}\delta^2/|v|^2$ vs $\delta$ for different
values of $u$ and $v$. }\label{asymp}
\end{figure}

The maximum probability of the system in the upper state is given
by $P_{\rm max}={\rm Max}_{\tau}[P_c]$. Figure \ref{Pmax} shows
the dependence of $P_{\rm max}$ on $\delta$. The solid lines are
analytical curves using (\ref{EqPmax}), and the dashed ones
represent the results of numerical calculations. While the two
curves coincide at small $\delta$, they soon deviate from each
other as $\delta$ increases. However, the overall behavior of the
curves, especially the asymptotic $P_{\rm max}\sim |v|^2/\delta^2$
dependence remains unchanged even at large $\delta$. To emphasize
on this aspect, we have plotted $P_{\rm max}\delta^2/|v|^2$ vs
$\delta$ in Fig.~\ref{asymp}, for different values of parameters.
All the curves overlap at large $\delta$ suggesting $P_{\rm max}
\sim |v|^2 / \delta^2 \sim (\Omega_R/\delta)^2$, in agreement with
(\ref{Panal}); the coefficient $\gamma$, however, is now a slow
function of the parameters, but still $O(1)$.

\section{Effect of pulse shape}

So far we have assumed that the microwave signal starts at
$\tau=0$, and continues forever. To perform a gate operation,
however, one needs to apply the Rabi signal for only a short
duration of time. In that respect, our calculation can only
describe hard pulses, in which the microwave switches on and off
abruptly. The probability $P_c$ then oscillates with the pulse
duration at the Rabi frequency. The maximum probability usually
happens in the case of a $\pi$-rotation, i.e when the probability
is maximally transferred to $|b\rangle$. A hard pulse, however, is
neither practical, nor the best pulse shape, as was indicated in
Ref.~\onlinecite{steffen}. Indeed, by using other types of pulses,
the probability of transition to the upper level, at the end of
the process, can be significantly reduced. Among a few pulse
shapes examined in \cite{steffen}, Gaussian pulses demonstrated
the most promise. To understand the role of pulse shaping, let us
compare the effect of a Gaussian pulse on the probability $P_c$,
with that of a hard pulse, for the case of a $\pi$-rotation.

To enforce a Gaussian envelope for the microwave signal V(t), we
write
%
%
%
\ba
 u(\tau)= \left\{\begin{array}{cc}
 ({a\Gamma/\tau_w}) e^{-(\tau-\tau_p/2)^2/2\tau_w^2}
 & \ \ {\rm for} \ 0<\tau<\tau_p \\
 0 & {\rm otherwise}
\end{array} \right., \nonumber
\ea
where $\tau_p$ and $\tau_w$ are the duration and width of the
pulse respectively, $\Gamma$ is the total angle of rotation in the
Bloch sphere (e.g.~$\Gamma=\pi$ for a $\pi$-rotation), and $a$ is
a normalization constant.

\begin{figure}[h]
\includegraphics[width=6cm]{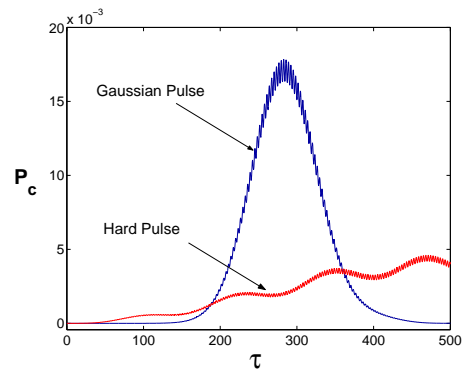}
\caption{Probability of the third level as a function of time for
a hard and a Gaussian $\pi$-pulse. }\label{Pct}
\end{figure}

Figure~\ref{Pct} shows the probability $P_c$ as a function of time
for a Gaussian and a hard pulse, both of which having the same
duration and resulting in a $\pi$-rotation ($\Gamma=\pi$) at the
end of the pulse. In our numerical calculation we take $v=u$,
$\delta=0.05$, $\tau_p=500$, $\tau_w=\tau_p/6$, and $a=0.398$.
These numbers correspond to the optimal pulse shape suggested in
\cite{steffen}. The maximum of $P_c$ for the Gaussian pulse,
happens slightly after the center of the pulse, while in the case
of the hard pulse, it occurs near the end. Although the maximum is
larger for the Gaussian pulse, the probability $P_{cf}$ at the end
of the process is much smaller. Orders of magnitude reduction of
the final probability can be achieved using such a technique.

\begin{figure}[h]
\includegraphics[width=6cm]{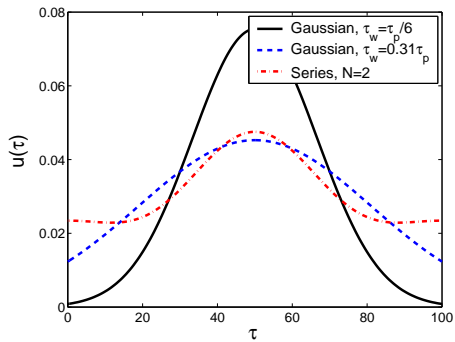}
\caption{Pulse shapes optimized for a $\pi$-rotation with
$\delta=0.1,\ \tau_p=100$. }\label{PShape}
\end{figure}

In Ref.~\onlinecite{steffen}, $\tau_w$ was fixed (to $\tau_p/6$ or
$\tau_p/4$) and $\tau_p$ was varied to minimize $P_{cf}$. A
$\tau_p \approx 8\pi/|\delta|$ was shown to provide the first
minimum with shortest duration. Alternatively, one can fix
$\tau_p$ and find a $\tau_w$ which gives minimum $P_{cf}$. This
may work better for shorter pulses. For example, for $\tau_p=100$,
$\delta=0.1$, and $v=u$, a Gaussian pulse with $\tau_w=\tau_p/6$
gives $P_{cf}=0.093$, while the minimum probability
$P_{cf}=0.0026$ is achieved at $\tau_w = 0.31 \tau_p$ and
$a=0.467$. Such a pulse shape starts and ends with jumps (see
Fig.~\ref{PShape}), but still gives smaller $P_c$ at the end of
the process.

\begin{figure}[h]
\includegraphics[width=6cm]{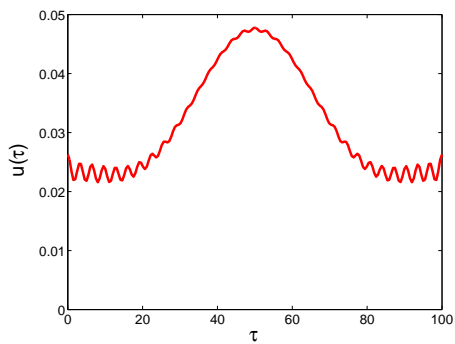}
\caption{Pulse shape of Eq.~(\ref{series}), optimized with $N=33$.
}\label{P40}
\end{figure}

A Gaussian pulse shape is not the optimal pulse shape for
minimizing $P_{cf}$. One can design other pulses with more free
parameters to achieve a smaller probability. To have some idea
about how small can $P_{cf}$ be made by appropriately shaping the
pulse, we defined an arbitrary pulse by the series
\be
 u(\tau) = (\Gamma/\tau_p)  \left[ 1 + \sum_{n=1}^N \lambda_n\cos
 (2\pi n \tau/\tau_p) \right]. \label{series}
\ee
Keeping only the first two terms in the series, (using the same
conditions as above: $\tau_p=100$, $\delta=0.1$, and $v=u$) one
can already reach a probability as small as $P_{cf}=1.2\times
10^{-5}$ with $\lambda_1=-0.3833$ and $\lambda_2=0.1293$ (see
Fig.~\ref{PShape}). With $N=33$ terms in the series, the
probability was reduced to $2.4 \times 10^{-6}$. The resulting
pulse shape, shown in Fig.~\ref{P40}, is complicated and may not
be useful experimentally. It should also be emphasized that with
the pulse shape of (\ref{series}), there is not a unique minimum
for $P_{cf}$. Depending on the starting point and the method of
minimization, one may fall into a local minimum with complicated
pulse shape.

Here, we only considered the case of $\Gamma=\pi$. For quantum
operations, other pulses may also be required. It is not just
enough to change the amplitude of the pulse, keeping its shape and
duration, to obtain optimized pulses with other $\Gamma$'s.
Indeed, for each type of operation, one needs to design a specific
pulse shape that provides minimum $P_{cf}$.

\section{Discussion}

In a practical quantum computer, the maximum number of operations
is limited by the decoherence time of the qubits as well as the
speed of operations. It is generally believed that if $\sim10^{4}$
operations can be performed within the decoherence time, quantum
computation can continue indefinitely with the help of quantum
error correction algorithms. A parameter that is commonly quoted
as a measure for the maximum number of operations is the quality
factor of the qubits, usually defined as
\be
 Q_\varphi={1\over2} \tau_\varphi \label{Q},
\ee
where $\tau_\varphi$ is the dephasing time of the qubit (in units
of $1/\omega_0$). $Q_\varphi$, however, is related to only one
type of single qubit operations, namely phase rotation. Other
necessary operations such as single qubit state flip or
multi-qubit gate operations are usually much slower. Even for the
phase rotation, the extent to which one can control the rotation,
i.e. change $E_{10}$, may be much smaller than the rotation
frequency itself.

The single qubit state flip can be performed using Rabi
oscillations \cite{martinis,vion,chiorescu} or non-adiabatic
evolution \cite{nakamura}. The latter is fast ($\approx
\omega_0$), but requires large anharmonicity to avoid unwanted
Landau-Zener transition to the upper states. Rabi oscillations, on
the other hand, are much slower, but can be used in small
anharmonicity systems. It is possible to define a quality factor
for the Rabi oscillations the same way as $Q_\varphi$ was defined
in (\ref{Q})
\be
 Q_R \equiv {1\over2} \Omega_R \tau_R \approx \Omega_R Q_\varphi \label{QR},
\ee
where $\tau_R$ is the Rabi decay time which is typically the same
order as $\tau_\varphi$.

In an ideal two level system, $\Omega_R$ is limited by the maximum
allowed amplitude of the microwave signal (restricted by RWA
and/or experimental limitations). Usually an $\Omega_R$ as large
as $0.1$ or even larger is conceivable. In practical systems,
especially those with small anharmonicity, however, increasing the
microwave power will cause transition to the upper states as we
discussed. Therefore $\Omega_R$ is limited by how much probability
of the upper levels can be tolerated. If we restrict $P_{\rm max}$
to $\sim 10^{-4}$, then (\ref{Panal}) gives $\Omega_R \sim 10^{-2}
\delta$. Therefore to achieve $\Omega_R\sim0.1$ ($Q_R\sim
0.1Q_\varphi$), we need a $\delta
> 10$. Such a large anharmonicity cannot be supported by many
qubit implementations (see below for a few examples).

Using a shaped (instead of hard) pulse can significantly reduce
the final $P_c$. To define a quality factor similar to (\ref{QR}),
we use the fact that in the case of a hard pulse, a $\pi$-rotation
is implemented when $\Omega_R=\pi/\tau_p$. We therefore define
\be
 Q_{\rm shaped}\equiv {1\over2} \left( {\pi \over \tau_p} \right)
 \tau_\varphi = \left( {\pi \over \tau_p} \right) Q_\varphi \label{Qshaped}.
\ee
Therefore a $Q_{\rm shaped}=0.1Q_\varphi$ requires a pulse with
duration $\tau_p=10\pi \approx 30$ for a $\pi$-rotation. It was
shown in \cite{steffen}, that a Gaussian pulse with
$\tau_w=\tau_p/6$ provides minimum $P_c$ with shortest time if
$\tau_p \approx 8\pi/|\delta|$. A quality factor of $0.1Q_\varphi$
is therefore achievable in a system with $\delta \approx 0.8$.
Other pulse shapes may provide better performance at smaller
$\delta$, as was discussed before. Below, we provide a few
examples among superconducting qubits.

In the current biased Josephson junction qubit of
Ref.~\onlinecite{martinis}, the energy differences are
$\omega_{10} \approx 6.9$ GHz and $\omega_{21} \approx 6.28$ GHz,
leading to $\delta \approx -0.09$.
%
%
Also, one can easily justify \cite{steffen} that
$|v|=\sqrt{2}|u|\sim |u|$, as expected. For a hard pulse,
requiring $P_{\rm max} \sim 10^{-4}$ and using (\ref{Panal}) (with
$\gamma \approx 1$), one finds $\Omega_R \sim 10^{-3}\omega_0$,
which is extremely slow. The quality factor $Q_R$ will also be
very small ($\sim 10^{-3}Q_\varphi$).
Aiming for a larger quality factor, one can make use of shaped
pulses. A Gaussian pulse with duration $\tau_p = 100$ ($Q_{\rm
shaped} \approx 0.03 Q_\varphi$) and with optimized width
($\tau_w=36.4$) gives $P_{cf}=0.0073$, which may not be small
enough. The pulse shape of Eq.~(\ref{series}), optimized with only
first two components ($\lambda_1=-0.2331,\ \lambda_2=0.2916$), on
the other hand, gives a probability as small as $P_{cf}= 1.6
\times 10^{-5}$, for the same pulse duration. It is not easy to
reach a small $P_{cf}$ with a shorter pulse.

In the charge-phase (quantronium) qubit of Ref.~\onlinecite{vion},
$\delta \approx 0.2$, $\Omega_R \sim 100$ MHz, and $\omega_0
\approx 16$ GHz. We therefore obtain $|u| \approx 0.0063$, and
with $|v|\sim|u|$, using (\ref{Panal}) we find $P_{\rm max}\sim
5\times 10^{-4}$ for a hard pulse, which is reasonably small. The
quality factor for the Rabi oscillation, however, is $Q_R \approx
150$ much smaller than $Q_\varphi =25000$ quoted in \cite{vion}.
Increasing the Rabi frequency will increase the probability
$P_{\rm max}$. With the help of a Gaussian pulse shape (with
optimal width $\tau_w=15.3$), a pulse duration of $\tau_p = 50$
(quality factor $Q_{\rm shaped} \approx 0.06 Q_\varphi$) is
achievable with $P_{cf}=0.0026$. Again, significant improvement in
the probability ($P_{cf}= 9.3\times 10^{-6}$) can be achieved
using Eq.~(\ref{series}), optimized keeping only two components in
the series ($\lambda_1=-0.4058,\ \lambda_2=0.1241$).

In practice, the shape of the pulse should be motivated
experimentally. For example, the jumps at the ends of the pulses
shown in Fig.~\ref{PShape} can only be realized approximately.
Such limitations should be considered as a constraint in the
optimization process. The minimization procedure may also be
preformed experimentally; trying different pulses with a few free
parameters and probing the transition probability to the upper
levels.

\section{Summary and conclusions}

We have performed analytical and numerical investigations of Rabi
oscillations in a three level system. We showed that the
probability $P_c$ of finding the system in the upper level
oscillates with the Rabi frequency $\Omega_R$. The maximum
probability $P_{\rm max}$ happens close to half a Rabi period. We
demonstrated that $P_{\rm max} \sim (\Omega_R/\delta)^2$, even
beyond RWA and when $\delta$ is large.

We also studied the effect of pulse shaping on $P_c$. We showed
that with an appropriate pulse shape, one can achieve small
probability $P_c$ at the end of the process, although in the
middle of the operation it may become large. The duration and
shape of the pulse can be optimized to obtain smallest $P_{cf}$ in
a shortest time. For each type of necessary operation, a specific
pulse shape should be designed. In any case, smallness of $\delta$
limits how short the pulse can be and therefore affects the speed
of qubit operations.

It is also necessary to take into account the effect of
decoherence on the studied phenomenon. In practice, however, only
a few Rabi oscillations happen during the operation. Thus, as long
as the decoherence time of the system is much longer than the Rabi
period, our conclusions remain valid even in the presence of
decoherence.

In this article, we only considered three levels. If the
anharmonicity of the system is very small, one needs to consider
more than three states. In Ref.~\cite{claudon}, $\sim 10$ states
were taken into account in the numerical simulations. Finally, we
should mention that having a multi-level, instead of two-level,
quantum system is not necessarily a disadvantage, as long as
coherent control of all the levels is possible. There have been
proposals to use multi-level systems for quantum computation
\cite{mls}.

\section*{Acknowledgment}

The author is grateful to A.J.~Berkley, A.~Maassen van den Brink,
A.Yu.~Smirnov, W.N.~Hardy, and A.M.~Zagoskin, for fruitful
conversations, and A.N.~Omelyanchouk for discussion and numerical
advice.

\end{document}